\documentstyle[twocolumn,aps]{revtex}
\input psfig.tex
\def\YBCO{$YBa_2 Cu_3 O_{7-\delta}$}
\def\YBCZO{$YBa_2 (Cu_{1-x}Zn_x)_3O_{7-\delta}$}
\def\ltsim{\vbox {\hbox{\lower 0.9\baselineskip \hbox{$<$}} \break
		 \hbox{\lower 0.2\baselineskip \hbox{$\sim$}} } }
\begin{document}
\draft

\twocolumn[\hsize\textwidth\columnwidth\hsize\csname %
@twocolumnfalse\endcsname

\title{
Theory of Thermal Conductivity in \YBCO
}

\author{
P.J. Hirschfeld$^a$ and W.O. Putikka$^b$
}

\address{
$^a$Dept. of Physics, Univ. of Florida, Gainesville, FL 32611, 
USA.\\
$^b$Dept. of Physics, Univ. of Cincinnati, Cincinnati, OH 45221-0011
}

\maketitle
\begin{abstract}
We calculate the electronic thermal conductivity in a d-wave 
superconductor, including both
the effect of impurity scattering and inelastic scattering by 
antiferromagnetic spin
fluctuations. 
We analyze existing experiments, particularly with regard to  
the question of the relative importance of electronic and phononic 
contributions to the heat current, and to the influence of 
disorder on low-temperature properties.  We find that phonons 
dominate
heat transport near $T_c$,  but  that electrons are responsible for 
most of
the peak observed in clean samples, in agreement with a recent 
analysis of Krishana et al.  In agreement with recent data 
on \YBCZO $~$, the peak position is found to vary nonmonotonically
with disorder.
\end{abstract}
]
{\it Introduction.}  Analysis of transport experiments in the 
superconducting
 state of the high-
temperature cuprate superconductors has already provided the most 
compelling evidence
for electronic pairing in these materials.  The collapse of the 
quasiparticle relaxation rate
below $T_c$, as observed in optical\cite{Nussetal,Romeroetal} and 
microwave\cite{Bonnetal} measurements, is not observed in classic 
superconductors,
and is most naturally interpreted in terms of a gapping of the spectral 
density of 
electronic excitations responsible for inelastic scattering just above 
$T_c$.  This collapse 
is now understood to be responsible, e.g., for the peak at 
intermediate temperatures in the
 microwave conductivity of 
$YBa_2Cu_3O_{6.95}$.\cite{Bonnetal}
\indent

Thermal conductivity measurements provide information 
on order parameter 
symmetry and quasiparticle
relaxation, and  have the advantage that they are bulk probes not subject to 
extrinsic surface effects
which have hampered the interpretation of the low-T 
microwave conductivity.\cite{Bonnetal}  They have the 
disadvantage that the electronic contribution 
to the heat current  must be separated from the phononic one.
As pointed out by Yu et al.,\cite{Yuetal}  the similarity between the 
microwave
conductivity peak and measurements of the thermal conductivity 
$\kappa (T)$ in
\YBCO ~ single crystals suggests that at least part of the thermal 
conductivity peak
should be due to the electronic thermal conductivity 
$\kappa_{el}(T)$, in contrast to earlier
analyses of this peak in terms of a phonon conductivity 
$\kappa_{ph}(T)$ 
alone.\cite{Peacoretal,Uherreview}  In the work of Yu et al., the 
phononic mean free path
$\ell_{ph}$ is assumed to vary only weakly with $T<T_c$, 
whereas in the Peacor et al.
approach, $\ell_{ph}$ is assumed to be dominated by phonon-
electron relaxation, leading
via the pair correlations in the electronic system to an exponential 
behavior below $T_c$.
In Ref. \cite{Peacoretal} it is assumed that 
$\kappa_{ph}\gg\kappa_{el}$ 
over the entire temperature range, whereas Yu et al. deduce 
$\kappa_{ph}(T_c) \simeq 2-3
\kappa_{el}(T_c)$ using the measured $\sigma_1(T_c)$ on 
similar quality crystals, and
assuming the 
Wiedemann-Franz
law $\kappa_{el}=L_0T\sigma_1$ with the free electron Lorenz number $L_0$.
\indent

In this paper, we adopt a theoretical model of electronic transport in 
a d-wave 
superconductor limited by impurity and spin fluctuation scattering 
which has proven
 successful in describing many of the systematics of microwave 
measurements, and apply it 
to calculate the electronic thermal conductivity.  We analyze 
experiments on  $Zn$- 
doped \YBCO $~$ to argue that a) phonons do in fact dominate heat 
conduction 
at $T_c$; b) a peak in $\kappa_{ph}$ does indeed occur at about 
20-25K; c) electronic 
conduction does nevertheless provide most of the peak in clean 
samples; and d) the
temperature at which the peak in $\kappa(T)$ 
occurs may vary 
nonmonotonically with disorder.  Conclusions a)-c) agree  with 
a recent thermal
Hall conduction measurement and analysis by Krishana et 
al.\cite{Krishanaetal}   Conclusion
d) agrees well with recent data on \YBCZO $~$ over a wide range of 
$Zn$ concentrations.\cite{Tailleferetal} 
We 
further focus on the very low-temperature behavior of the thermal 
conductivity
in the d-wave model, discussed recently in considerable detail by 
Graf et al.\cite{Grafetal}
In particular, we analyze the disorder and phase shift dependence of 
the ``universal''
linear-T term in $\kappa$ predicted recently by several 
groups.\cite{SunMaki,Grafetal,NH}  Since at  very low 
temperatures 
the phonon mean free path has saturated, this 
contribution is a direct reflection
of electronic correlations and relaxation.
\indent

{\it Electronic thermal conductivity.}
The electronic thermal conductivity $\kappa$  for an unconventional 
superconductor\cite{ph1,srink,ph2,otherthermcond}
is evaluated using a Kubo formula
for the heat-current response as in the original treatment
for an s-wave superconductor.\cite{amb}   In this letter we 
particularly wish to study
the $d_{x^2-y^2}$ pair state thought to provide a good description 
of  optimally doped
\YBCO\cite{scalapinoPR,pinesreview}.  For simplicity we work in 
what follows with
the approximate order parameter $\Delta_k = \Delta_0 \cos (2 \phi )$ 
over a circular Fermi surface
to describe ab plane transport.  The impurity-averaged matrix 
(Nambu) electron 
propagator in such a state is given by 
\begin{eqnarray}
\underline{g}({\bf k},\omega) 
 =  
\frac{\tilde{\omega} \underline{\tau}
^0 + \xi_{\bf k}
\underline{\tau}^3 + i{\Delta}_{\bf k} \underline{\sigma}^2
\underline{\tau}^1  } 
{\tilde{\omega}^2 
- 
\xi_{\bf k}^2 - 
|\Delta_{\bf k}|^2}
\end{eqnarray}
\noindent
where 
$\underline{\sigma}^i$ and $\underline{\tau}^i$ are the Pauli 
matrices in spin and particle-hole space, respectively.
 Here, we have already 
exploited the assumed particle-hole symmetry of the normal state, as 
well as
the symmetries  of the gap functions which lead to  vanishing 
renormalizations for 
both the order parameter and the single-particle energies.  
In this case, only self-energy contibutions
to the frequency $\omega$, namely 
$\tilde{\omega}= \omega - \Sigma_0$ need 
to
be included.\cite{ph2}
The self-energy $\Sigma_0$ due to the elastic impurity scattering is
treated in a self-consistent $t$-matrix approximation and
is given by
$\Sigma_0 
 =  
\Gamma G_0/(c^2-G_0^2),$
\noindent
where $\Gamma=n_i n/(\pi N_0)$ is the normal state unitarity 
limit scattering rate  depending on the concentration of defects 
$n_i$,
the electron density $n$, and the density of states at the Fermi level 
$N_0$.
The quantity $c\equiv \cot \delta_0$ 
parameterizes the scattering strength of an individual
impurity through the s-wave phase shift $\delta_0$.  In this work 
we 
consider only near-unitarity limit scattering $c\simeq 0$ since it is clear
that weak scattering will lead to a weak temperature dependence 
inconsistent
with experiment for the states in question.
The integrated propagator is 
$
G_0 
 =  
(1/2\pi N_0) \sum_{\bf k}  \mbox{Tr} \{\underline{
\tau}^0\underline{g}({\bf k},\omega)\}.
$
\noindent
The equation for the self-energies 
are then solved self-consistently together with the 
gap equation
$
\Delta_{\bf k} 
 =  
-T \sum_{\omega_n}\sum_{\bf k'} V_{\bf kk'}(1/2)
\mbox{Tr}\{\underline{\tau}^1 \underline{g}
({\bf k'},\omega_n)\}~.
$
\indent

The above approximation is insufficient to describe transport at 
temperatures close
to $T_c$, where inelastic scattering is known to dominate.  As in 
Refs.\cite{HPS,HPSPRB},
we adopt a model of scattering by antiferromagnetic spin 
fluctuations  based on an
RPA treatment of the Hubbard model with parameters chosen to 
reproduce normal state NMR
and resistivity data in \YBCO.\cite{Quinlanetal}  The relaxation rate 
due to spin fluctuations $1/\tau_{in}$
is quasilinear in temperature above $T_c$, and falls as $\sim T^3$ 
in the superconducting state
due to a) a crossover to Fermi liquid behavior below a spin 
fluctuation scale, and b), an additional
factor of $T$ due to the restriction of relevant quasiparticle momenta 
to the vicinity of the d-wave
order parameter nodes.\cite{Quinlanetal} 
 To include inelastic scattering in the model in a crude way, we make 
the replacement 
$\Sigma_0\rightarrow \Sigma_0-i/2\tau_{in}$.  Although we have 
adopted a particular microscopic model, we emphasize that the two 
features a) and b) noted
may be common to many models of inelastic relaxation, which
will then lead to qualitatively similar results.
\indent

The bare heat current response is now given by a convolution of
the Green's function $g$ with itself at zero external frequency and 
wave vector
weighted with the bare heat current vertex 
$\omega {\bf v}_{F{\bf k}} \underline{\tau^3}$.\cite{amb}
Impurity scattering vertex corrections to current-current correlation 
functions  at $q=0$
have been shown to vanish identically for even
parity states ($\Delta_{\bf k} = \Delta_{-{\bf k}}$).\cite{ph2} For 
the diagonal thermal conductivity tensor one obtains
\begin{eqnarray}
\frac{\kappa_{el}^i(T)/T}{\kappa^{N,i}_{el}(T_c)/T_c} 
=
\frac{6}{\pi^2}
\int_0^{\infty}{d\omega}\left(\frac{\omega}{T}
\right)^2 \left( {-\partial f\over \partial \omega}\right) 
K_i(\omega , T)\\[10pt]
\label{K}
K_i(\omega , T) 
 =  
\frac{\Gamma_{tot}(T_c)}{\tilde{\omega}^\prime
\tilde{\omega}^{\prime\prime}}
{\rm Re}\left\langle \hat{\bf k}_i^2 
\cdot \frac{\tilde{
\omega}^2 + |\tilde{\omega}|^2 - 2|\Delta_{\bf 
k}|^2}{\sqrt{\tilde{\omega}^2
- |\Delta_{\bf k}|^2}}\right\rangle_{\hat{\bf k}},
\end{eqnarray}
\noindent
where ${\tilde\omega}^\prime$ and 
${\tilde\omega}^{\prime\prime}$ are the
real and imaginary parts of $\tilde \omega$, $f$ is the Fermi 
function, and $\Gamma_{tot}\equiv\Gamma +1/2\tau_{in}$ is
the total quasiparticle scattering rate.  We have numerically 
evaluated Eqs. (2-3), and show results in Figure 1.  In the clean limit, 
the combination of the relaxation rate collapsing  with decreasing 
temperature due to gapping of the spin fluctuation
spectral density and the rapidly decreasing number of quasiparticles 
at low $T$ leads to a peak in
the thermal conductivity very similar to that found for the electrical 
conductivity.\cite{Bonnetal,HPS}  As impurities are added, the 
collapse of the
inelastic scattering rate is cut off at progressively higher and higher 
energy scales, such that the
peak moves to higher temperatures and simultaneously weakens.  
A preliminary report on these findings is contained in Ref. \cite{Miamipaper}.
\vskip -.9cm
\begin{figure}[p]
\leavevmode\centering\psfig{file=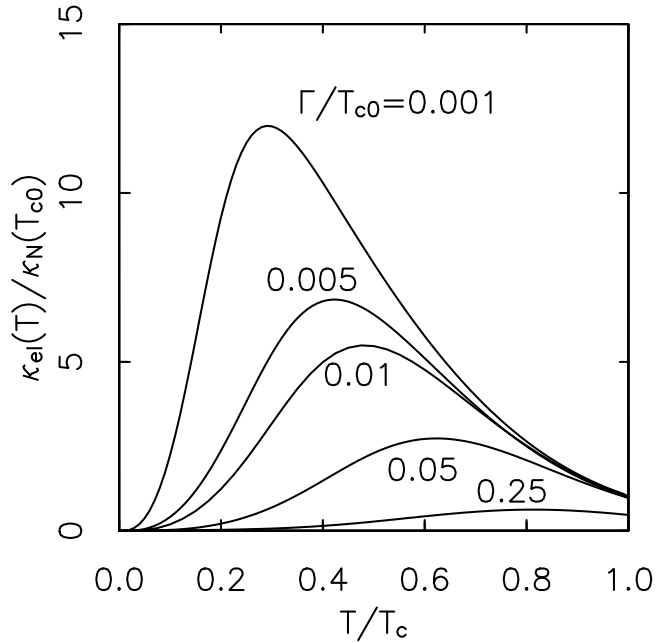,width=\columnwidth}
\caption{Normalized electronic thermal conductivity 
$\kappa/\kappa_N$ for
normalized impurity scattering rate $\Gamma/T_{c0}$ = 
0.001,0.005,0.01,0.05, and 0.25, $\Delta_0 /T_c=3$ and $1/\tau_{in}(T_c)
=T_c$}
\end{figure}
\vskip -.6cm
\indent

{\it Phonon thermal conductivity.}  
Were the phonon 
and electron thermal 
conductivities comparable at $T_c$, the electronic peak in clean 
samples would, according
to Figure 1, be very large,
leading to ratios $\kappa(T_{peak})/\kappa(T_c)$ much larger than the 
observed range of about 
1.8-2.4.\cite{Yuetal,Krishanaetal,Miamipaper,Tailleferetal}  
We therefore
expect  on theoretical grounds
that the phononic
conductivity 
at $T_c$ is several times larger than the electronic 
conductivity.\cite{Miamipaper}
This conclusion was reached independently by Krishana et al.,\cite{Krishanaetal}
who 
performed a semiclassical
Boltzmann type analysis of their measurements of the diagonal and 
off-diagonal components of
the tensor $\kappa_{ij}$ in a magnetic field, assuming as the origin 
of the off-diagonal terms
skew-scattering of quasiparticles off the vortex lattice.  The diagonal 
phonon conductivity
 $\kappa_{ph}$ was then extracted from the data under the 
assumption that only electrons 
were skew scattered.
Here we adopt the $\kappa_{ph}$  as shown in Figure 2 following the
analysis of Ref. \cite{Krishanaetal}, using a value of the transverse
scattering cross section such that $\kappa (T_{peak}) - \kappa (T_c)$
is primarily of electronic origin.\cite{Krishanaetal}
  We use the form of $\kappa_{ph}$ shown
for all calculations
at any impurity concentrations, thereby neglecting the weak effect of 
point defects on the long
wavelength phonon mean free path, $\tau_{point}^{-1} \sim 
\omega^4  $.
The overall scale of $\kappa_{ph}$ is set by its value at $T_c$, 
which is found to be roughly
seven times $\kappa_{el}(T_c)$.
\begin{figure}[p]
\leavevmode\centering\psfig{file=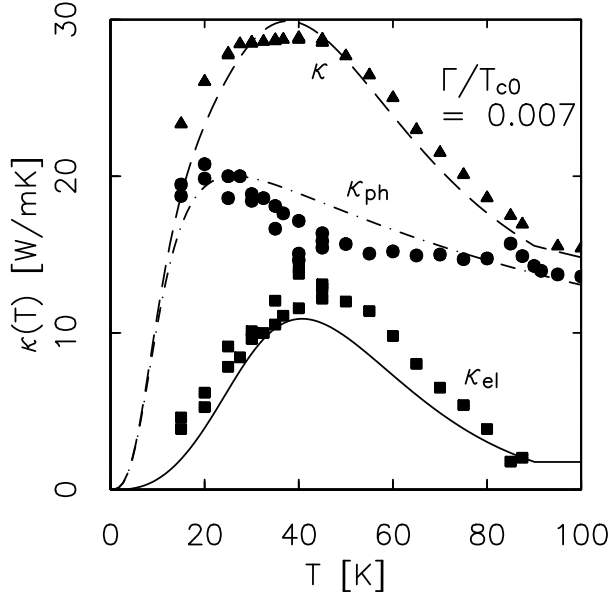,width=3in}
\caption{Solid line:  electronic thermal conductivity $\kappa_{el}$;
dash-dotted line: phononic thermal conductivity $\kappa_{ph}$;
dashed line: $\kappa$ vs. $T/T_c$ for
$\Gamma/T_{c0}=0.007$. Data  are from Ref. 7. }
\end{figure}
\vskip -.2cm
\indent
{\it Zn substitution: high temperatures.}
In Figure 2, we plot 
the theoretical $\kappa(T)$  obtained by adding $\kappa_{ph} (T)$,
 determined as above, to $\kappa_{el} (T)$, with disorder parameter $\Gamma$
chosen to give rough agreement with the peak height and position of
the nominally clean sample of Ref. \cite{Krishanaetal}.
We note that the quasiparticle mean 
free path extracted from nominally pure crystal data in Ref.\cite{Krishanaetal},
is actually similar to the mean free path in the 0.15\% Zn sample 
of Ref. \cite{Bonnetal},
making the assignment  $\Gamma /T_c =0.007$ consistent with earlier 
analysis.\cite{HPSPRB}
In Figure 3 we now show results for the total thermal conductivity with
systematic $Zn$ doping compared
to data of Ref.\cite{Tailleferetal} on twinned crystals of \YBCZO.
Note that the impurity scattering
parameters $\Gamma$ for the various curves were chosen to reflect
$Zn$ concentrations of
0.1\%,0.2\%,0.7\%,1.7\% using the  identification of a contribution
to the scattering rate
$\Gamma/T_{c0} =0.12$ per 1\% $Zn$.
This is a factor of 2 larger than
the contribution
extracted from comparisons with microwave data in Ref.\cite{HPSPRB},
but consistent with the error bars cited in that work.
We have also  added in all cases a residual
scattering rate of $\Gamma /T_c=0.007$ consistent with  
the nominally pure sample as in Figure 2, apparently representative
of oxygen defects in the near-optimally doped samples.
With this assignment, there are no further free parameters in the theory.
\indent
We note first that the position of the peak in 
temperature is nonmonotonic in both the data and the theory.  In the
current theory this occurs 
because, while the electronic peak initially dominates and
moves upward in $T$ with disorder, the phonon peak at 20-25K 
eventually becomes more
important as the scale of $\kappa_{el}$ is reduced by disorder.  We 
expect that continued $Zn$ doping will lead to  saturation and a
peak position fixed at $20-25K$ when $\kappa_{el}$ disappears, leaving
the (roughly impurity independent) $\kappa_{ph}$ of Fig. 2.
\vskip -.3cm
\begin{figure}[p]
\leavevmode\centering\psfig{file=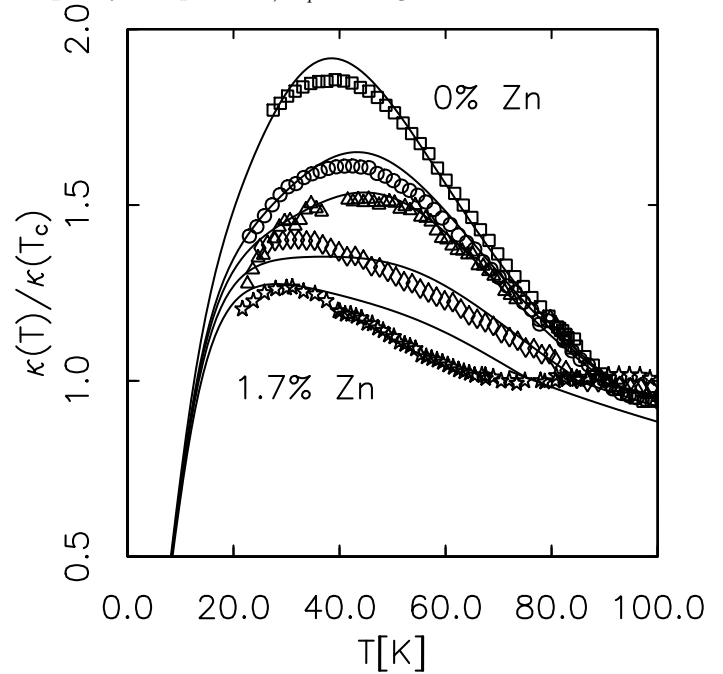,width=3.4in}
\caption{Total thermal conductivity $\kappa$ vs. $T[K]$. Symbols
are data from Ref. 8,  $Zn$ concentration 0.0\%,
0.1\%,0.2\%,0.7\%,1.7\%.  Solid lines: theoretical $\kappa$
for $\Gamma/T_{c0}=0.007+0.12(\%Zn)$.}
\end{figure}
\indent
\vskip -.4cm
{\it Wiedemann-Franz and anisotropy ratios.}  
Measured values of the
ratio $\kappa(T_c)/T_c\sigma(T_c)$ are typically $1.0-1.5 \times
10^{-7}W-\Omega/K^2$ .  If we take 
$R\equiv \kappa_{ph}(T_c)/
\kappa_{el}(T_c)= 7$ as above, we find for the Lorenz number 
$L = \kappa_{el}(T_c)/T\sigma(T_c)= 1.2-2.0 \times 10^{-8}W-\Omega/K^2$. 
Given that electron
correlations can change this ratio substantially,
this is 
quite close to the
free electron value of $L_0=2.44\times 10^{-8}W-\Omega/K^2$.  
The measured anisotropy ratio in the plane, $\kappa_b 
(T_c)/\kappa_a(T_c)$ is 1.2-1.3 for
clean untwinned crystals.\cite{Yuetal,Tailleferetal}  If we assume that all 
anisotropy arises from the electronic component,
and using $R\simeq 7$ once again, we find 
$\kappa_{b,el}(T_c)/\kappa_{a,el}(T_c) = 2.6-3.4.$ 
This is close to the measured electrical conductivity ratio of 
$\sigma_b/\sigma_a \simeq 2.4$
reported by Zhang et al.\cite{Zhangetal}.
\indent

{\it Zn substitution; low temperatures.}
Direct information on  electronic properties may be obtained by working
at very low temperatures, such that the phonon contribution, which should
fall as $T^3$ when the mean free path saturates due to boundary scattering,
is negligible.  In the resonant scattering limit considered here,
a significant linear-$T$ electronic contribution $\kappa_{el}\simeq
aT$ should dominate the phonon conductivity in a d-wave
superconductor.  The thermal conductivity
in this limit is ``universal'' in the sense that the prefactor $a$
is independent of the impurity scattering rate to leading 
order,\cite{SunMaki,Grafetal,NH} in 
analogy to the limiting $T=0, \omega\rightarrow
0$ conductivity $\sigma_{00}\equiv (ne^2)/(m\pi\Delta_0)$.\cite{PAL} 
As pointed out by Graf et al.
the Wiedemann-Franz law is obeyed exactly at $T=0$ in the clean limit, i.e. 
$a\rightarrow L_0\sigma_{00}$.
\vskip -2.5cm
\begin{figure}[p]
\leavevmode\centering\psfig{file=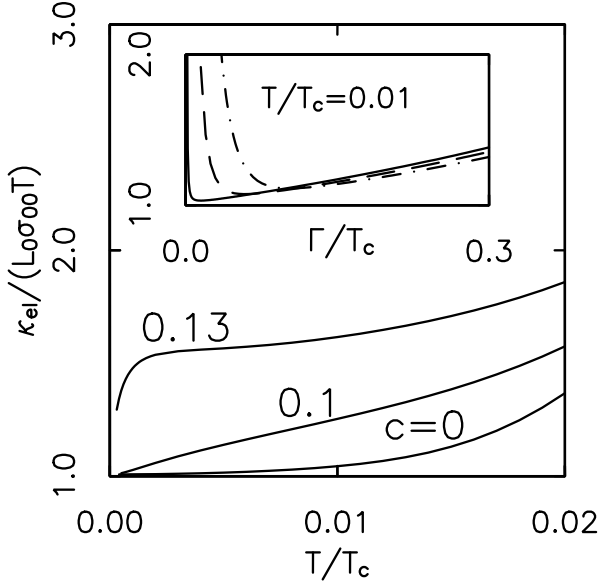,width=4in}
\caption{Normalized electronic thermal conductivity 
$\kappa_{el}/L_0\sigma_{00}T vs. T/T_c$ for 
varying electronic scattering phase shifts, $c=0,0.1,0.13$, $\Gamma/T_c=0.007$. Insert:
$\kappa_{el}/L_0\sigma_{00}T$ vs. $\Gamma/T_c$ for fixed 
$T=0.01T_c$ for $c=0$ (solid line); $c=0.1$ (dashed line); and $c=0.3$
(dashed-dotted line).
}
\end{figure}
While $\sigma_{00}$ is
difficult to measure due to its small size and to possible extrinsic
contributions alluded to above, the linear term in $\kappa (T)$ should
be clearly visible, and its magnitude and impurity dependence sensitive
tests of the ``dirty d-wave'' model.
\indent 

The exact expression for the limiting value of $\kappa_{el}/T$ for
$T\ll T_c$ is\cite{SunMaki,Grafetal}
$a=L_0 \sigma_{00} k^\prime {\bf E} (k^\prime )$,
where $k^\prime=\sqrt{\Delta_0 / (\Delta_0^2 + \gamma^2)}$,
$\bf E$ is a complete elliptic integral of the 2nd kind, and
$\gamma=-{\rm Im}\Sigma_0(\omega=0)$.
In the unitarity limit, $c=0$, the $\kappa_{el}=aT$ relation holds
over a temperature range $T\ltsim \sqrt{\Gamma \Delta_0}$, whereas
away from this point the range of validity quickly vanishes.  
A quasilinear behavior may nonetheless be observed; for example,
in the opposite limit limit, $c\gg 1$, we have 
$\kappa_{el}(T)/T\simeq [\kappa_{el} (T_c)/T_c](2\Gamma\tau(T_c))^{-1}$
above an exponentially small crossover scale $\sim \Delta_0 {\rm exp}
-\Delta_0(1+c^2)/\Gamma$.
To illustrate
the possible range of behavior we plot in Fig. 4
$\kappa_{el}/T$ at   $T/T_c =0.01$ for various values of $c$ close to the
unitarity limit,  and
as a function of  $Zn$ concentration.
The main point we wish to make here is that it is possible that small
deviations from the unitarity limit may lead
at low T 
to behavior quite different
from that predicted for  unitarity limit scattering.
\indent

{\it Conclusions.}  We have shown that ``dirty d-wave theory'' provides
a good account of the systematic behavior of the thermal conductivity
in \YBCZO.  In particular, it allows for a simple understanding of
the size and position  of the intermediate temperature
peak as a function of disorder.  The analysis is consistent with
earlier studies of the microwave conductivity, enabling 
semiquantitative predictions.  On this basis we
have provided strong evidence for a phononic conductivity significantly
larger than its electronic counterpart near $T_c$.  The electronic
component is nevertheless responsible for the peak in clean samples,
but disappears with per cent level Zn doping.  Finally, we have discussed
the low-T limiting behavior of $\kappa$, and pointed out that small
deviations from the unitarity scattering limit can give rise to quite
nonuniversal results.
\vskip .4cm
\indent
{\it Acknowledgements.}  The authors gratefully acknowledge 
extensive
discussions with J. Crow, P. Henning, N.P. Ong, S. Quinlan,
and L. Taillefer, and are particularly grateful to the last
for providing data prior to publication.
W.O.P. was partially supported by NSF-DMR-9357199.


\vskip -.5cm
\begin{references}
\vspace*{ -1.5cm}
\bibitem{Nussetal} M.C. Nuss et al.  Phys. Rev. Lett. 66, 3305 (1991).

\bibitem{Romeroetal} D.B.  Romero et al. Phys. Rev. Lett. 68, 1590 (1992).

\bibitem{Bonnetal} D.A. Bonn et al., Phys. Rev. B47, 11214 (1993).

\bibitem{Yuetal} R.C. Yu et al., Phys. Rev. Lett. 69, 1431 (1992);
J. Supercond. 8, 449 (1995).

\bibitem{Peacoretal} S.D. Peacor et al,  Phys. Rev. B44, 9508 (1991).

\bibitem{Uherreview} C. Uher, in {\it Physical Properties of 
High Temperature Superconductors}, D.M. Ginsburg, ed. (World Scientific,
Singapore, 1992), V. 3.

\bibitem{Krishanaetal} K. Krishana et al. Phys. Rev. Lett. 75, 3529 (1995).
N.P. Ong, private communication.

\bibitem{Tailleferetal} 
L. Taillefer, R. Gagnon, B. Ellman, B. Lussier, S. Pu, R. Besserling, K.
Behnia, and H. Aubin, to be published.
 
\bibitem{Grafetal} M.J. Graf et al., Phys. Rev. B 53, 15147(1996).

\bibitem{SunMaki} Y.  Sun and K. Maki Europhys. Lett. 32, 355 (1995).

\bibitem{NH} M. Norman and P.J. Hirschfeld Phys. Rev. B53, 5706 (1996).  

\bibitem{ph1}
P.\,J.\,Hirschfeld et al.,
Sol.\, St.\, Commun. \, {\bf 59}, 111 (1986)
 
\bibitem{srink}
 S.\,Schmitt-Rink et al., Phys.\, Rev.\, Lett. {\bf 57},
 2575 (1986)
  
\bibitem{ph2}P.\, J.\,Hirschfeld et al.Phys.\,Rev.\,B\,{\bf 37},83 (1988).

\bibitem{otherthermcond} 
B.\,Arfi et al., Phys.\,Rev.\, B {\bf 39}, 8959 (1989);
H. Monien et al.,Sol. St. Commun. {\bf 61},581 (1987).
	     
\bibitem{amb}
V.\,Ambegaokar and  A.\,Griffin, Phys.\,Rev.\,{\bf 137}, A1151 
(1965)

\bibitem{scalapinoPR} D.J. Scalapino,
Physics Reports 250,329 (1995).

\bibitem{pinesreview} D. Pines and P. Monthoux, J. Phys.
Chem. Solids 56, 1651 (1995).


\bibitem{HPS} P. J. Hirschfeld,  W.O. Putikka, and D. Scalapino, 
Phys. Rev. Lett. 71, 3705
(1993).

\bibitem{HPSPRB} P.J. Hirschfeld, W.O. Putikka, and D. Scalapino,
Phys. Rev. B. 50, 10250 (1994).

\bibitem{Quinlanetal}
S. M. Quinlan et al., Phys. Rev. B {\bf 49}, 1470 (1994).

\bibitem{Zhangetal} K. Zhang et al., Phys. Rev. Lett. 73, 2484 (1994).

\bibitem{PAL} P.A. Lee, Phys. Rev. Lett. 71 ,1887 (1993).


\bibitem{Miamipaper} P. Henning et al., J. Supercond. 8, 453 (1995).  
\end{references}
\end{document}